# FEM MODELLING TECHNIQUES FOR SIMULATION OF 3D CONCRETE PRINTING

Gieljan VANTYGHEM[1], Ticho OOMS[1], Wouter DE CORTE[1]

1. Department of Structural Engineering and Building Materials, Ghent University, Ghent, Belgium

Corresponding author email: Gieljan.Vantyghem@UGent.be

## Abstract

Three-dimensional concrete printing (3DCP) has gained a lot of popularity in recent years. According to many, 3DCP is set to revolutionize the construction industry: yielding unparalleled aesthetics, better quality control, lower cost, and a reduction of the construction time. In this paper, two finite element method (FEM) strategies are presented for simulating such 3D concrete printing processes. The aim of these models is to predict the structural behaviour during printing, while the concrete is still fresh, and estimate the optimal print speed and maximum overhang angle to avoid print failures. Both FE analyses involve solving multiple static implicit steps where sets of finite elements are added stepwise until failure. The main difference between the two methods is in the discretization of the 3D model. The first method uses voxelization to approximate the 3D shape, while the second approach starts from defining the toolpath and constructs finite elements by sweeping them along the path. A case study is presented to evaluate the effectiveness of both strategies. Both models are in good agreement with each other, and a comparable structural response is obtained. The model's limitations and future challenges are also discussed. Ultimately, the paper demonstrates how FEM-based models can effectively simulate complex prints and could give recommendations with regards to a better print strategy. These suggestions can be related to the maximum printing speed and overhang angle, but also the optimal layer height and thickness, the specific choice of the infill pattern, or by extension the mixture design. When print failures can be avoided, this methodology could save time, resources and overall cost. Future work will focus on the validation of these numerical models and comparing them to experimental data. The developed Grasshopper plug-in '*CobraPrint*' can be downloaded from the following website: www.food4rhino.com/app/Concre3DLab

## 1. Introduction

Three-dimensional printing of concrete (3DCP) is a challenging but promising new production technique that has received a lot of attention in recent years. It is also one of the current focus points within 'Construction 4.0', which is a term used to refer to the digitalization of the construction industry (Craveiro et al. 2019). A 3DCP set-up usually comprises a concrete (or cementitious) mixing and pumping system and a mechanism for precise positioning control (often a gantry system or an industrial 6-axis robotic arm). A growing number of research groups is experimenting with building their own 3DCP set-up, causing a fast-evolving technical environment. Nevertheless, improvements in the field can be made on many different levels. For example, some groups are more focused on robotics and purifying the mechanical aspects involved in concrete printing, while others concentrate more on the material science (e.g. improving the buildability, extrudability, and pumpability of the material, or finding more sustainable material mixtures). At present, most distinction (i.e. outreach) is acquired by those who focus on large-scale experiments such as printing a one-story building in a single day (Hager,

Golonka, and Putanowicz 2016) or constructing a full-scale 3D-printed concrete bridge (Salet et al. 2018). However, printing at large scale requires rapid hardening of the material (without premature cracking) and having a good understanding of the complete printing process. For this, adequate fresh mechanical and thixotropic behaviour of the concrete is needed (i.e. the thixotropic behaviour ensures shape stability of an individual layer; and high yield stress materials with fast stiffness evolution over time provide the overall stability during collective layers accumulating). Nevertheless, working with high yield stress materials may result in weak interfaces (Panda et al. 2019). While it may seem like some have found a 'secret recipe', it remains very hard to predict whether a design is 'printable' or not. The trending methodology therefore consists of intensive trial and error procedures, which leads to an incredible waste of resources. Process simulation of 3D concrete printing aims to resolve this problem. By simulating the print process in a virtual world, the chance of success can be increased. The main economic benefit is therefore that it can limit the number of costly physical experiments. Nevertheless, the simulation remains largely dependent on the accuracy of the material model.

In this paper, two novel strategies for simulating concrete printing processes are presented that are based on the finite element method (FEM). Of course, several other models already exist that can estimate the buildability performance of a 3DCP element. For example, Suiker (2018) proposed a mechanistic model to analyse and optimize the printing of straight wall structures. The model distinguishes between two failure mechanisms: elastic buckling of the global structure and plastic collapse at the bottom layer. Results demonstrated a good agreement with the experimental buckling response, but the model is not suited for free-form shapes. Alternatively, Roussel (2018) presented a set of analytical equations that describe the rheological requirements for printable concrete structures. Here, aspects like layer interface strength, strength-based stability of the first layer, and overall buckling stability are discussed. Although these give rough estimates of the buildability, the application of these equations is again not valid for all (especially curved) shapes. In contrast, Wolfs, Bos, and Salet (2018) were the first to propose a FE model to study the mechanical behaviour of concrete in the fresh state. In this model, a virtual copy of the design is imported, and a static/implicit solver is used to make predictions. The model is first divided into printable layers, which are added in a stepwise fashion on top of each other, until completion (or failure). The time-dependent material properties are implemented in a Mohr-Coulomb failure criterion and linear stress-strain behaviour up to failure. Using improved material characterization methods, as described in Wolfs, Bos, and Salet (2019), the model can make reasonable predictions (i.e. a 15% overprediction of the total number of printable layers compared to the experimental results for a 5 m long wall structure). However, in the paper no examples are presented to use this method for free-form designs or generally more complex geometries already printed in practice. Additionally, the paper leaves rooms for further numerical improvements.

In this paper, we built upon the work by Wolfs, Bos, and Salet (2018) and propose two improved strategies for simulating concrete printing processes using numerical methods. In the next section, the conceptual build-up of the two methods is presented, and finally, a simulation of a complex shape design is presented.

## 2. Method

The model proposed by Wolfs, Bos, and Salet (2018) is used as the starting point of this paper. The software package SIMULIA Abaqus is used as the main FE solver, and Rhinoceros © (Grasshopper) are used to generate the input files of the simulation model. There are two main differences compared to Wolfs' model. On the one hand, adjustments are made to the way in which the FE mesh is created and secondly how the input file is written. On the other hand, the Abaqus parameters and numerical settings are tuned for better structural response. For example, an automatic stabilizer is used in both methods to better analyse (i.e. visualize) the buckling behaviour. Finally, two custom Grasshopper plug-ins were developed, each of which can produce a complete Abaqus (calculation) file as a result from a random input geometry. The two distinct methods are named: (i) VoxelPrint, and (ii) CobraPrint, as derived from their corresponding Grasshopper component names. In the following sections, the general build-up of each of the two methods is discussed: first describing their general concepts, whereafter, certain model specific limitations and individual (dis)advantages are discussed.

*2.1. VoxelPrint*

The first method uses voxels to represent the 3D-printed structure. Voxels are the 3-dimensional equivalent of pixels. So, like rasterizing a vector graphic (shape) into a raster image (a series of pixels), a 3D model or shape can be 'voxelized' into a set of 3D unit cubes (voxels). The term 'voxelization' is used in this paper to describe the process of transforming a random 3D shape into a group of voxels (Fig. 1). In Grasshopper, a 3D discrete space is built from many of these small cubes, and if the centre of such cube is in close proximity of the 3D model, it will be activated. This way, any kind of 3D shape can be voxelized, regardless of its complexity (e.g., self-intersection and layer contact), as such, the generation of the FE mesh is straightforward; a direct link can be made between the generated voxels and the FE mesh (eight-node continuum elements). To improve the computation time, a thick line drawing algorithm (IBM Corporation 1978) was also implemented but requires a predefined printing path as input. However, this can be easily created by using a (contour) slicer for the input geometry. At the same time, the printing path can determine the sequential (or stepwise) activation of adjacent print segments. The main advantage of the method is that only one Abaqus part needs to be created, and no interaction, tie or contact constraints need to be defined between different segments containing the (finite) elements. A disadvantage of the method is that, when required, no advanced contact properties can be used, e.g. to model cold joints; so-called weakened adhesive behaviour between layers. Therefore, the model is only valid for prints with small time gaps between concurrent layers.

Constructing the Abaqus input file is as simple as extracting the base coordinates of the voxels and determining the corresponding node coordinates. The mesh elements and elements sets can be determined from the voxel indices and the toolpath sequence. In the Abaqus input file, the mesh elements (voxels) are first *deactivated* in the initial step, using the Model Change function; and step-wisely *reactivated* in each subsequent analysis step. If the toolpath is self-intersecting or if contact between layers occurs during printing, this is automatically apprehended by the model. In the case of self-intersections, the voxels that are already active, will remain active, and in the case of contacting elements, the adjacent voxels will 'come to life', and the predefined connection is established. Similar to the work by Wolfs, Bos, and Salet (2018), time-dependent material behaviour is assigned by adding field variables to the element sets when they become activated. In this method, the field variables are constructed such that their value increases linearly over time (step count) using the Amplitude function in Abaqus.

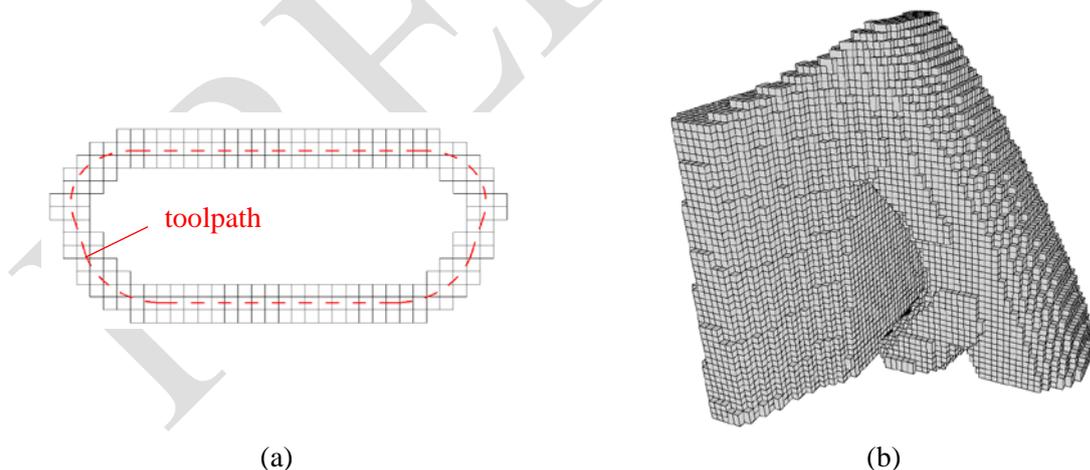

(a)          (b)

**Figure 1.** Voxelization of a predefined printing part using a thick line drawing algorithm (a) and example of a fully-'voxelized' 3D model by the Grasshopper component (b).

*2.2. CobraPrint*

In this approach, a structured mesh is generated by sweeping a cross-section of the printed concrete layer along the print path (Fig. 2) This mesh discretization is realized by a custom Grasshopper code; by first dividing the print path into several segments, and projecting vertices tangential to the path and in the z-direction. By careful parametrization, the approximate size of the mesh elements along the path can be used as an input. The final FE model contains a number of meshed layers, divided into segments, which are, similar to the previous method, activated sequentially to simulate the printing process. Using the Abaqus Model Change command, the new sets of elements are activated step by step. In this method, the transformation from the 3D-printed structure to the FE mesh is much more accurate and even allows for bevels on the layer's edges. However, attention must be paid to the minimum curvature of the print path in order to avoid intersecting neighbouring elements. Also, layer contact is much more difficult to model, as all intersecting elements must be determined beforehand. The name CobraPrint was derived from the snake-like build-up of the model.

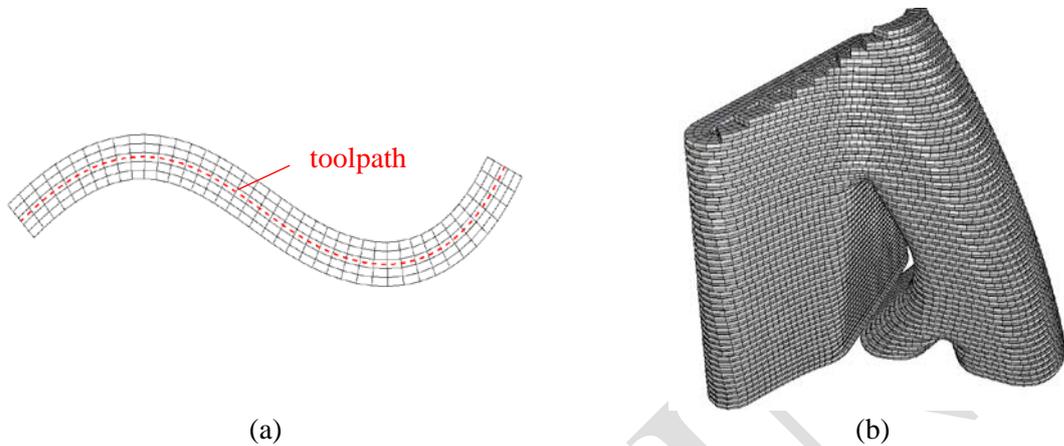

(a)  (b)
**Figure 2.** Mesh generation process of CobraPrint along a predefined printing path (a), and an example of a 3D meshed model using CobraPrint (b).

*2.3. Extensions to the model of Wolfs et al.*

In extension to the concept that was first proposed by Wolfs et al., the two presented methods are provided with the following numerical modifications: First, numerical damping is introduced in the form of an automatic stabilization mechanism. This mechanism is used in Abaqus to stabilize unstable quasi-static problems through the addition of volume-proportional damping and is here used to analyse the buckling response of the 3D-printed structure during failure. Without this stabilization, the Abaqus solver would prematurely abort the analysis when the structure starts to collapse. As such, the failure mode cannot be analysed in detail. The damping factors that were used need to be quite small as to not influence the structural behaviour. Secondly, it also aids to stabilize the model - specifically for complex designs - where partially-collapsing regions can occur. After local failure, the structure can stabilize again (e.g. a new contact between a neighbouring wall segment is established). Alternatively, such class of unstable problems can also be solved dynamically (explicit) or with the aid of (artificial) material damping (e.g. using dashpots). However, these methods were not (yet) adopted in this paper. Some smaller numerical modifications include using eight-node brick elements (C3D8) instead of four-node tetrahedral elements (C3D4) because the latter are too stiff and not ideal for use in structural calculations, unless their number is drastically increased. Furthermore, other improvements with regards to ease of work: the automatic generation of the complete Abaqus code and the straightforward parametrization in the Grasshopper plug-ins also add value.

*2.4. Comparison between both methods*

Before discussing the results from the case study, the (dis)advantages of both methods are briefly presented. A first advantage of VoxelPrint is the straightforward implementation of the discretizer. VoxelPrint is much simpler than CobraPrint, and as such, less prone to bugs. Contact is automatically established, and the method does not need special attention when simulating complex shapes, like minimum curvature control or having multiple closed curves per layer. A main disadvantage of VoxelPrint in contrast to CobraPrint is its low mesh approximation and no opportunity for advanced contact interactions. Examples of shapes that are not printable with (the current version of) CobraPrint are square and triangle shapes with sharp edges. Lastly, VoxelPrint has a much faster calculation time compared to CobraPrint.

*2.5. Material model*

The time-dependent material properties were extracted from the research papers by Wolfs et al. on early-age and triaxial compression testing of 3D-printed concrete. We used the following properties, presented in Table 1. The essential material parameters to set up the Mohr-Coulomb material model are the material density $\rho$, the Young's modulus $E$ function of time ($t$), Poisson's ratio $v$, cohesion $c$ function of time ($t$), the angle of internal friction $\varphi$ and the dilatancy angle $\psi$. As the focus of this paper is on a comparison between numerical methods, a detailed explanation of these material parameters falls beyond the scope of this study and the reader is redirected to the above-mentioned papers.

**Table 1**. The Mohr-Coulomb time-dependent material properties as used in the numerical models.

| Mix design | $\rho$ [kg/m³] | $E(t)$ [kPa] | $v$ [-] | $c(t)$ [kPa] | $\varphi$ [°] | $\psi$ [°] |
| --- | --- | --- | --- | --- | --- | --- |
| Weber 145-2 | 2100 | 1.705 t + 39.48 | 0.24 | 0.0636 t + 2.60 | 20 | 13 |

## 3. Case study

The case study that is presented in this paper was taken from the design of a 3D-printed concrete girder designed by topology optimization, described in Vantyghem et al. (2020). In this paper, the digital design and manufacturing of a post-tensioned concrete girder is demonstrated (Fig. 3). The design of the girder was optimized using advanced topology optimization algorithms in which not only the concrete shape, but also the tendon profile was optimised (Amir and Shakour 2018). Additionally, the optimised shape was then realized using 3D concrete printing techniques, classified as formwork printing. While designing the segments, the maximum height of each segment had to be determined in order to avoid print failure caused by plastic collapse or elastic buckling. Additionally, very large overhang angles were existent in some of the segments, which had to be checked for their printability. During the initial experiments, large deformations (and buckling of the longest straight wall) were observed (Fig. 4). Therefore, adjustments to the design had to be made. One of these adjustments was the addition of an internal support structure, positioned in the mid-section to avoid buckling of that wall. Artefacts of this internal support are still visible in the final structure (Fig. 3b). Secondly, it became clear that the steep overhang angle at some points was going to remain a problem. For this, the following research question was urgently put on the agenda: Can we simulate the behaviour of fresh 3D printed concrete? Additionally, it would be interesting to know certain design limitations for a preselected printing mixture and to be able to answer the following questions: "What is the maximum printing speed to print this specific model?" and "What is the maximum overhang angle this material can tolerate?" In the end, the simulation models were not ready before the realisation of this project, and many elements were printed using sand as a support.

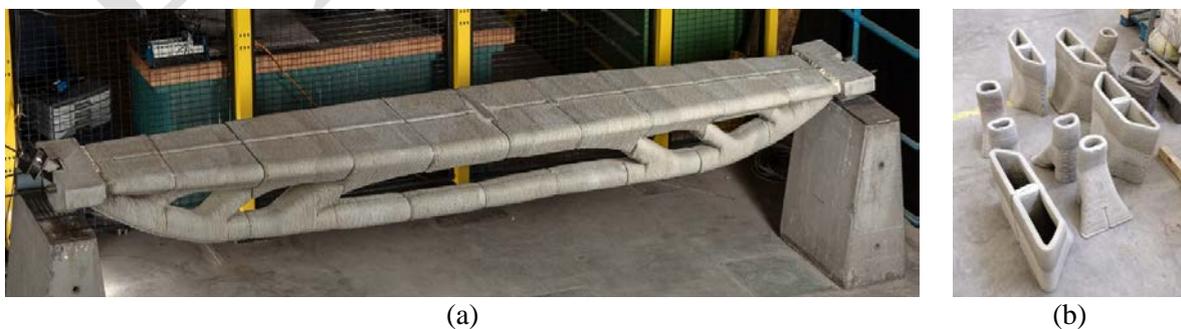

(a)      (b)

**Figure 3**. Completed girder (image courtesy of Vertico) (a), and the 3D-printed segments before assembly showing the internal support (b) taken from (Vantyghem et al. 2020).

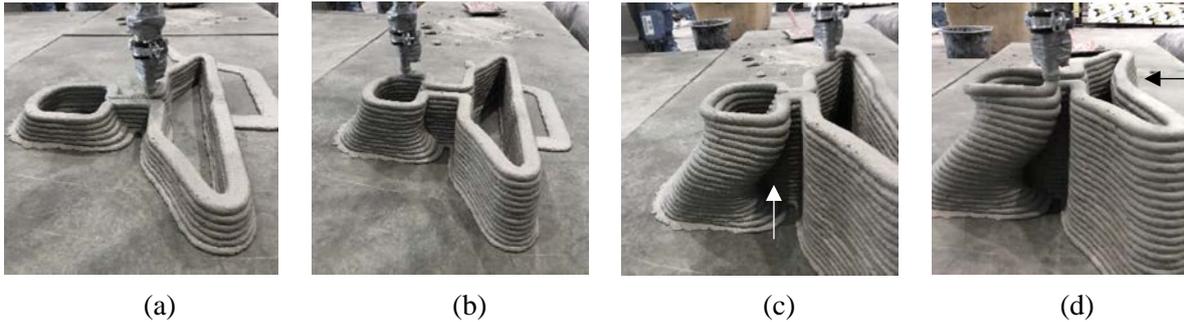
(a) (b) (c) (d)
**Figure 4**. Image of an early experiment showcasing large deformations of the fresh concrete during printing.

For this paper, we will simulate the segment closest to the support of the bridge as this segment contains the most challenging modelling difficulties: large overhangs, small curvature in the corners, multiple toolpaths per layer, open and closed layers, etc. A detailed view of this segment's CAD geometry, toolpath and realisation is presented in Figure 5. Important to note, is that the toolpath of the actual realization was optimized in order to improve the buildability (Fig 5b). However, in this paper the printing sequence follows a layer-wise approach where all the bottom layers are printed first, then the second layers, and so on. To use such a traditional print strategy, an automated start/stop valve (i.e. to close off the extrusion during travel paths) would need to be implemented. In the future, this additional print path optimization could of course be included in the FE model builder as well.

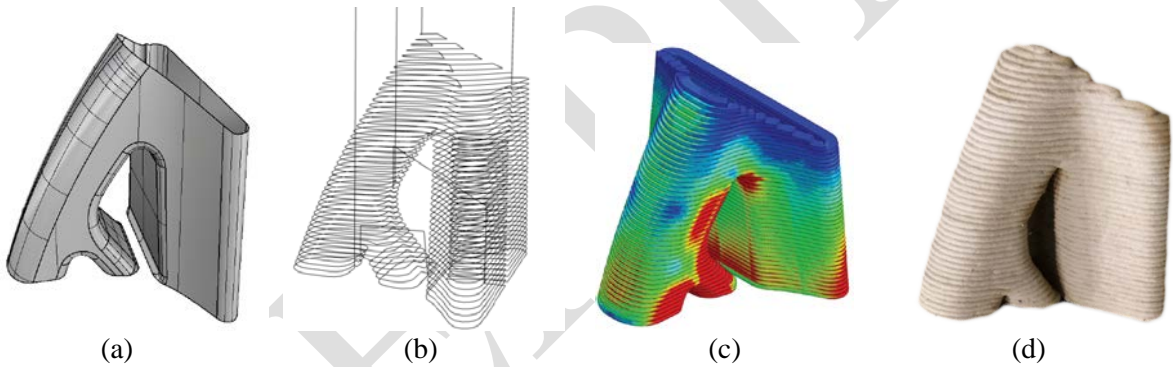
(a) (b) (c) (d)
**Figure 5.** From design to print: (a) Rhino (polysurface) model, (b) toolpath generation using Grasshopper contours, (c) process simulation (d) actual realisation.

The following paragraphs will now discuss the results of the numerical study proposed in the previous section. The first study simulates the end block being printed at a speed of 80 mm/s (Fig. 6).

From the result of both models, it appears that the print is not be able to complete because of 'very' early failure. The problem arises within one of the steeper overhangs of the model. The numerical model produced by CobraPrint fails around the start of the $10^{th}$ layer, while the model that uses the voxel method can reach 12 layers. In order to compare both models, we have plotted the deformations of the $8^{th}$ layer – right after its completion. Figure 6a and 6c present the deformations from CobraPrint, while Figure 6b and 6d present the results from VoxelPrint. The scale of the legend is set equal and shows displacements up to 10 mm. In all Figures 6-7, the non-scaled deformations are shown. As can be observed, a comparable structural response is obtained by both models, with deformations that are in the same order of magnitude. Both models fail by the collapse of the inclined wall; no plastic strains are present. In the actual realisation, sand was poured in at this location to support the steep overhang.

In order to improve the buildability of the material, the print speed was decreased allowing more time for the fresh concrete to harden. The next set of results shows the structural response for a print with a speed of $1/10^{th}$ the original speed (Fig. 7). This is of course mainly hypothetical, as printing at 8 mm/s would be very slow. Nevertheless, the results show that even at this speed, the complete segment cannot be printed. Now, the print overcomes the steep overhang, but fails at the 'straight wall' section. Here, local buckling arises, exactly where - for the actual realisation - the internal wall is providing support. Nevertheless, both models fail at the $19^{th}$ layer, illustrating excellent resemblances.

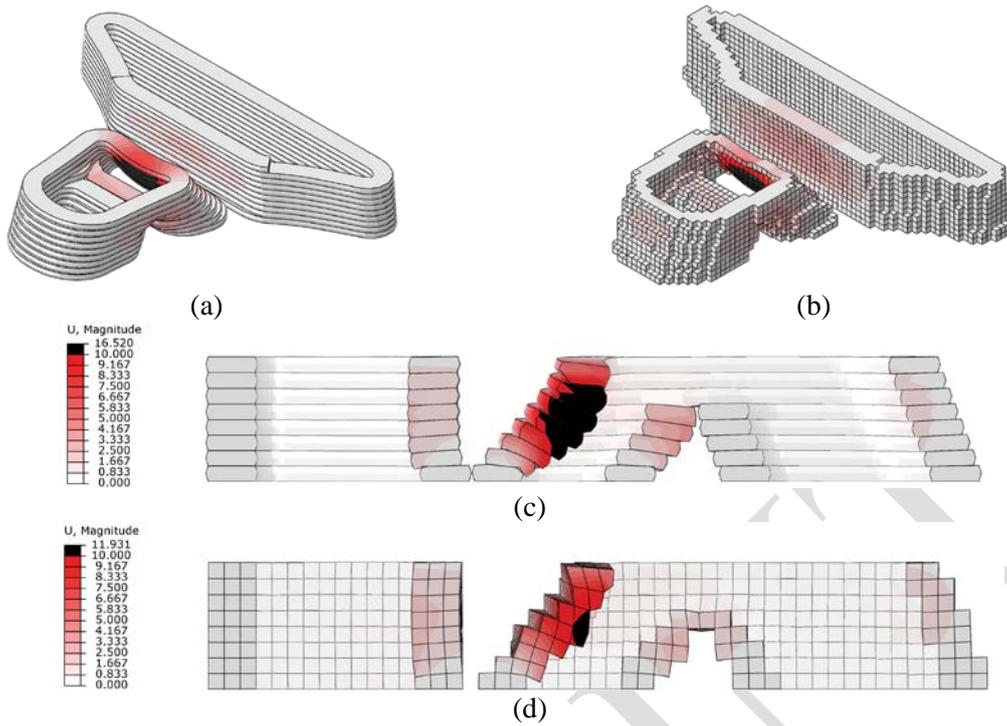

**Figure 6.** Results from the numerical simulation for a print speed of 80 mm/s: (a,c) CobraPrint, (b,d) VoxelPrint.

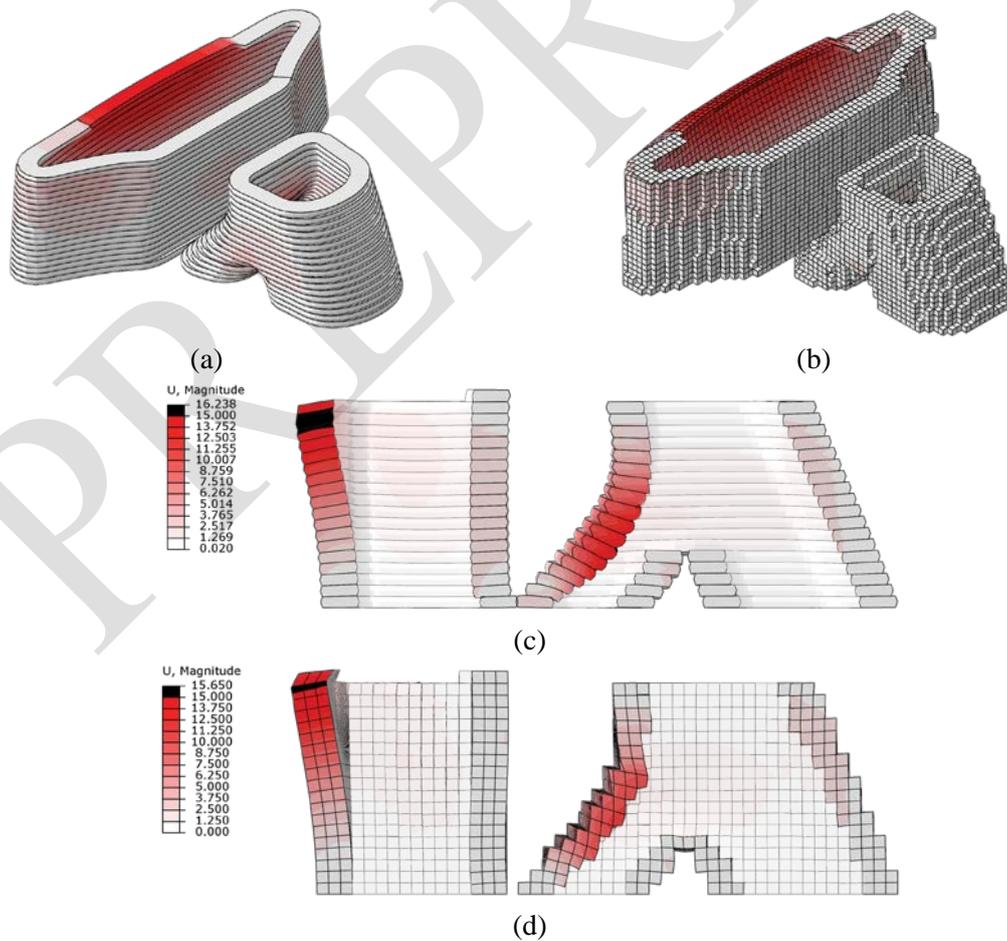

**Figure 7.** Results from the numerical simulation showing the non-scaled deformations in the model one layer before failure occurs. The print speed is 8 mm/s: (a,c) CobraPrint, (b,d) VoxelPrint.

# 4. Conclusions and future work

This paper has presented two extended FEM-based strategies for simulating 3D concrete printing. The developed scripts were explained and discussed, and their well-functioning was demonstrated by studying a rather complex case study. A comparison between both methods was made and their potential to provide quick preliminary data is clearly demonstrated. VoxelPrint is the fastest method that provides also the easiest implementation but has its limitations. For example, it could only be used for prints with small time gaps, as interlayer interaction properties could not be included. Finally, both simulation techniques could help construct design rules, and link digital design with the physical production process. It can be estimated that the use of such simulation models will become standard practice for validating of the success rate of a complex 3DCP design.

The focus of this paper has been on the numerical methodology, rather than a verification of the models by comparing them to a physical experiment. This aspect is still crucially lacking but is currently being investigated by the authors. Secondly, the phenomenon of elastic buckling of straight wall structures is also described in Wolfs and Suiker (2019). Several closed-form expressions are presented to calculate the critical wall buckling length of rectangular and straight walls. In order to verify our simulation models, the parametric model that was developed by Wolfs and Suiker could also be applied here. However, no such comparison was made at the current moment.

## Acknowledgements

This research project was supported by Ghent University.

## References

Amir, Oded, and Emad Shakour. 2018. "Simultaneous Shape and Topology Optimization of Prestressed Concrete Beams." *Structural and Multidisciplinary Optimization* 57 (5): 1831–43. https://doi.org/10.1007/s00158-017-1855-5.

Craveiro, Flávio, José Pinto Duarte, Helena Bartolo, and Paulo Jorge Bartolo. 2019. "Additive Manufacturing as an Enabling Technology for Digital Construction: A Perspective on Construction 4.0." *Automation in Construction*. Elsevier B.V. https://doi.org/10.1016/j.autcon.2019.03.011.

Hager, Izabela, Anna Golonka, and Roman Putanowicz. 2016. "3D Printing of Buildings and Building Components as the Future of Sustainable Construction?" In *Procedia Engineering*, 151:292–99. Elsevier Ltd. https://doi.org/10.1016/j.proeng.2016.07.357.

IBM Corporation. 1978. "IBM Technical Disclosure Bulletin 20 (May): 5358–5366. http://homepages.enterprise.net/murphy/thickline/index.html

Panda, Biranchi, Nisar Ahamed Noor Mohamed, Suvash Chandra Paul, G. V.P.Bhagath Singh, Ming Jen Tan, and Branko Šavija. 2019. "The Effect of Material Fresh Properties and Process Parameters on Buildability and Interlayer Adhesion of 3D Printed Concrete." *Materials* 12 (13). https://doi.org/10.3390/ma12132149.

Roussel, Nicolas. 2018. "Rheological Requirements for Printable Concretes." *Cement and Concrete Research*. Elsevier Ltd. https://doi.org/10.1016/j.cemconres.2018.04.005.

Salet, Theo A. M., Zeeshan Y. Ahmed, Freek P. Bos, and Hans L. M. Laagland. 2018. "Design of a 3D Printed Concrete Bridge by Testing." *Virtual and Physical Prototyping* 13 (3): 222–36. https://doi.org/10.1080/17452759.2018.1476064.

Suiker, A. S.J. 2018. "Mechanical Performance of Wall Structures in 3D Printing Processes: Theory, Design Tools and Experiments." *International Journal of Mechanical Sciences* 137 (March): 145–70. https://doi.org/10.1016/j.ijmecsci.2018.01.010.

Vantyghem, Gieljan, Wouter de Corte, Emad Shakour, and Oded Amir. 2020. "3D Printing of a Post-Tensioned Concrete Girder Designed by Topology Optimization." *Automation in Construction* 112 (April): 103084. https://doi.org/10.1016/j.autcon.2020.103084.

Wolfs, R. J.M., F. P. Bos, and T. A.M. Salet. 2018. "Early Age Mechanical Behaviour of 3D Printed Concrete: Numerical Modelling and Experimental Testing." *Cement and Concrete Research* 106 (April): 103–16. https://doi.org/10.1016/j.cemconres.2018.02.001.

Wolfs, R. J.M., F. P. Bos, and T. A.M. Salet. 2019. "Triaxial Compression Testing on Early Age Concrete for Numerical Analysis of 3D Concrete Printing." *Cement and Concrete Composites* 104 (November). https://doi.org/10.1016/j.cemconcomp.2019.103344.

Wolfs, R. J.M., and A. S.J. Suiker. 2019. "Structural Failure during Extrusion-Based 3D Printing Processes." *International Journal of Advanced Manufacturing Technology* 104 (1–4): 565–84. https://doi.org/10.1007/s00170-019-03844-6.